\documentclass[twocolumn,prd,superscriptaddress,preprintnumbers,amsmath,amssymb,nofootinbib]{revtex4-1}

\usepackage{subfigure}
\usepackage{amsmath,amssymb}
\usepackage{graphicx}
\usepackage{rotating}
\usepackage{bm}
\usepackage{color}

\def\di{\displaystyle}

\def\bg{\begin{eqnarray}\begin{array}{rcl}\displaystyle}
\def\eg{\end{array} &\di    &\di   \end{eqnarray}}
\def\bm#1{\begin{eqnarray}\begin{array}{#1}\di}
\def\bmo#1{\begin{eqnarray*}\begin{array}{#1}\di}
\def\bml#1#2{\begin{eqnarray}\begin{array}{#1}\label{#2}\di}
\def\bgo{\begin{eqnarray*}\begin{array}{rcl}\displaystyle}
\def\ego{\end{array} &\di    &\di \nonumber  \end{eqnarray*}}

\def\btensor#1#2{\renew\left#1\begin{array}{#2}\di}
\def\brtensor#1#2#3{\ren#3\left#1\begin{array}{#2}}
\def\botensor#1#2{\renew\left#1\begin{array}{#2}}
\def\etensor#1{\end{array}\right#1}

\def\eq#1{(\ref{#1})}
\def\Eq#1{Eq.~(\ref{#1})}

\def\id{1\!\mbox{l}}

\def\s0#1#2{\mbox{\small{$ \frac{#1}{#2} $}}}
\def\0#1#2{\frac{#1}{#2}}

\date{\today}

\def\ren#1{\renewcommand{\arraystretch}{#1}}

\def\renew{\renewcommand{\arraystretch}{1}}

\definecolor{blue}{rgb}{0,0,1}

\definecolor{green}{rgb}{0,1,0}

\definecolor{red}{rgb}{1,0,0}

\newcommand{\Tr}{\mathrm{Tr}}

\newcommand{\tr}{\mathrm{tr}}

\newcommand{\be}{\begin{eqnarray}}
\newcommand{\ee}{\end{eqnarray}}

\newcommand{\Gk}{\Gamma_k}

\newcommand{\Nc}{N_{\text{c}}}
\newcommand{\lqcd}{\Lambda_{\text{QCD}}}
\newcommand{\pat}{\partial_t}
\newcommand{\Eqref}[1]{Eq.~\eqref{#1}}

\usepackage{bbm}

\begin{document}

\title{Gluon condensation and scaling exponents for the propagators in
  Yang-Mills theory}

\vspace{.8cm}

\vspace{.6cm}

\author{Astrid Eichhorn} \author{Holger Gies}
\affiliation{Theoretisch-Physikalisches Institut,
  Friedrich-Schiller-Universit\"at Jena, D-07743 Jena, Germany}
\author{Jan M.~Pawlowski} \affiliation{Institut f\"ur Theoretische
  Physik, University of Heidelberg, Philosophenweg 16, 62910
  Heidelberg, Germany} \affiliation{ExtreMe Matter Institute EMMI, GSI
  Helmholtzzentrum f\"ur Schwerionenforschung, Planckstr. 1, 64291
  Darmstadt, Germany}

\begin{abstract}
  We investigate the infrared (strong-coupling) regime of
  SU($N$)-Yang-Mills theory on a self-dual background. We present an
  evaluation of the full effective potential for the field strength
  invariant $F_{\mu \nu}F^{\mu \nu}$ from non-perturbative gauge
  correlation functions and find a non-trivial minimum corresponding
  to the existence of a dimension four gluon condensate in the
  vacuum. We also relate the infrared asymptotic form of the $\beta$
  function of the running background-gauge coupling to the asymptotic
  behavior of Landau-gauge gluon and ghost propagators. Consistency
  between both gauges in the infrared imposes a new upper bound on the
  infrared exponents of the propagators. For the \emph{scaling
    solution}, this bound reads $\kappa_c < 23/38$ which,
  together with Zwanziger's horizon condition $\kappa_c> 1/2$,
  defines a rather narrow window for this critical exponent. Current
  estimates from functional methods indeed satisfy these bounds.
\end{abstract}

\maketitle

\section{Introduction}

The understanding of low-energy QCD has continuously been advancing
over the recent decades. Since a variety of methods, gauges and
pictures have been used to investigate QCD in detail, progress has
never been linear but many scenarios have been developed abreast.
Particularly for the challenging problem of confinement, a number of
successful scenarios exist in parallel. Contemporary research
therefore has to clarify whether these scenarios mutually support,
exclude or simply coexist with one another.  In this work, we pursue
this question for two scenarios of pure Yang-Mills gauge theory based
on two different gauges: the Landau gauge and the background gauge.

{In the Landau gauge, information about the strongly coupled sector
  and confinement is drawn from low-order correlation functions which
  have been studied with functional methods such as Dyson-Schwinger
  equations (DSE) \cite{von
    Smekal:1997is,Lerche:2002ep,Fischer:2002hn,Fischer:2006vf,Fischer:2008uz}, the
  functional Renormalization Group (FRG)
  \cite{FRG,Pawlowski:2003hq,Fischer:2004uk,
    Fischer:2006vf,Fischer:2008uz}, and stochastic quantization
  \cite{Zwanziger,Pawlowski:2009iv}; as well as with lattice gauge
  theory \cite{Pawlowski:2009iv,lattice}.} Many results support the
Kugo-Ojima, or Gribov-Zwanziger confinement scenarios
\cite{Kugo:1979gm,Gribov:1977wm,Zwanziger:1993dh} where the
impossibility of colored asymptotic states imposes conditions on the
infrared (IR) behavior of ghost and gluon propagators (see section
\ref{selfdual}). Reviews on the different approaches and numerical
results can be found in \cite{Fischer:2008uz,vonSmekal:2008ws,DSE_reviews};
 for further work see \cite{Cornwall:1981zr,Zwanziger:2009je}.

The background gauge is particularly useful for computing the full
quantum effective action in an expansion in terms of gauge-invariant
operators. As the quantum effective action governs the dynamics of
field expectation values, information about confinement is contained
in the resulting nonlinear and nonlocal quantum equations of
motion. Simple scenarios for the full effective action such as the
leading-log model \cite{Adler:1981as,Lehmann:1983bq,Dittrich:1996cd}
or more sophisticated dielectric confinement models
\cite{Fishbane:1986rg,Chanfray:1990fi,Hosek:1989up} indeed find
confining potentials for the solution of the field equations for
static color sources.

Even though the Landau gauge and the background gauge have formally
much in common, results from one gauge generally cannot easily be
transferred to the other. The background gauge reduces to covariant
Lorenz-type gauges including the Landau gauge in the limit of zero
background. However, most results for the background gauge are
obtained for zero fluctuation field instead, so that direct
Landau-gauge information is lost. In turn, the full
  background-field dependence in the background gauge cannot be
  reconstructed solely from Landau-gauge correlation functions. For
  special cases, however, such a reconstruction in certain parametric
  limits is indeed possible. This has recently been shown for the case
  of the Polyakov loop \cite{Braun:2007bx}, where Landau-gauge
  propagators have been used to determine a variant of the background
  Polyakov-loop potential. These results compare well with a related
  study in Polyakov-gauge, \cite{Marhauser:2008fz}.

The resulting connection between the two gauges work in both ways
\cite{Braun:2007bx}: given the Landau-gauge propagators, an estimate
of the deconfinement phase transition temperature has been computed in
excellent agreement with lattice simulations. The other way round, a
confining Polyakov-loop potential in the background gauge imposes a
new confinement criterion on gluon and ghost propagators in the Landau
gauge. This criterion is indeed found to be satisfied by these 
propagators.

In the present work, we further explore the connection between the two gauges,
concentrating on the running coupling and the effective potential for
the square of the Yang-Mills field-strength.  In particular we focus on
the interrelation between the asymptotic behavior of the gluon and ghost
propagator for low momenta and the $\beta$ function of the background
running coupling. Consistency of the latter with a strongly interacting
  infrared regime already imposes severe restrictions on the scaling exponents
  of the propagators.  Furthermore, we report on a non-perturbative
calculation of the full effective potential without any polynomial
truncation. This allows to search for a non-trivial minimum, indicating
the condensation of gluons in the vacuum.

This work is structured as follows: In Sect.~\ref{background}, we introduce a
non-perturbative Renormalization Group equation, the Wetterich equation for
the case of non-zero background and elaborate on the extension of Landau-gauge
propagators to the background gauge. We then specialize to the case of a
self-dual background in Sect.~\ref{selfdual}, which allows to evaluate the
asymptotic form of the $\beta$ function of the running coupling. This can be
related to the asymptotic behavior of the propagators, see
Sect.~\ref{betafunction}). Finally, we report on our results for a numerical
evaluation of the full effective potential in Sec.~\ref{effpot}. Conclusions and
discussion are summarized in Sect.~\ref{conclusion}.

\section{Background-field flow}\label{background}

Information about correlation functions and effective potentials is
encoded in the effective action $\Gamma$, being the generating
functional of 1PI Green's functions. Nonperturbative access to the
effective action is given by the functional Renormalization Group (RG)
in which the full $\Gamma$ is constructed successively by integrating
out quantum fluctuations momentum shell by momentum shell. For a
generic quantum field theory, this procedure leads to a flow equation
for a scale-dependent effective action $\Gamma_k$. Here, $k$ denotes a
momentum-scale above which all quantum fluctuations have been
integrated out. The dependence of $\Gamma_k$ on this momentum scale is
determined by the Wetterich equation \cite{Wetterich:1993yh}, being an
exact functional differential equation. Reviews on this topic can be
found in \cite{Reviews}, for gauge theories see
\cite{gaugereviews,Pawlowski:2005xe}.  The application of the
background-field method to flow equations in gravity is reviewed in
e.g. \cite{gravityreviews}.

In gauge theories, momentum-shell integrations in continuum quantum field
theory can most conveniently be performed in gauge-fixed formulations. An
effective action constructed from only gauge-invariant building blocks can
nevertheless be obtained by use of the background-field formalism
\cite{Abbott:1980hw,Reuter:1993kw}. Here the full gauge field
$\mathcal{A}_\mu$ is split into a background $A_{\mu}$ and a fluctuation field
$a_{\mu}$ according to $\mathcal{A}_\mu = A_\mu+a_\mu$. The gauge of the
fluctuations is then fixed with respect to the background field, whereas the
whole action remains invariant under an auxiliary gauge transformation of the
full gauge field $\mathcal{A}_\mu$ and the background field $A_{\mu}$. Let us
stress that this transformation is not the physical, but just an
auxiliary gauge transformation. It allows, however, to retain invariance of the standard
effective action $\Gamma[A]=\Gamma[a=0,A]$ under physical gauge
transformations. Within the functional RG setting $a=0$ is allowed only once
all fluctuations have been integrated out, hence the price to be paid for the
construction of a gauge-invariant effective action is its dependence on two
gauge fields, \cite{Pawlowski:2001df,Litim:2002xm}.

The Wetterich flow equation in background-field formalism reads
\cite{Reuter:1993kw,Reuter:1994zn}: 
\begin{equation}
 \partial_t \Gamma_k[a, {A}]= \frac{1}{2}{\rm STr}
\left(\Gamma_k^{(2,0)}[a, {A}]+R_k\right)^{-1}\partial_t R_k, 
\label{eq:flow}
\end{equation}
where $\Gamma_k^{(n,m)}[a, {A}]= \frac{\delta^n}{(\delta
  a)^n}\frac{\delta^m}{(\delta {A})^m}\Gamma_k[a, {A}]$ and $ \pat \equiv
 k\frac{d}{dk}$. The action $\Gamma_k$
is a functional which interpolates between the microscopic action $S_\Lambda$
at the UV cutoff $\Lambda$ and the full quantum effective action $\Gamma$,
i.e., $\Gamma_{k\to\Lambda}\to S_\Lambda$ and $\Gamma_{k\to 0}\to
\Gamma$. The solution to the flow equation \eqref{eq:flow} provides an RG
trajectory of action functionals $\Gamma_k$ that interconnects the microscopic
action with the full effective action. The quantity $R_k$ is a regulator
function depending on an infrared (IR) cutoff $k$ that suppresses propagation
of momenta smaller than $k$. The trace ($\rm STr$) runs over all internal indices, momenta
and field components, i.e., gluon and ghost degrees of freedom, including a negative sign for the ghosts. 

As we will eventually be
interested in the background-field action
\begin{equation}
{\Gamma}_k [ A] = \Gamma_k[a=0, A],
\end{equation}
simplifications arise from the fact that only the fluctuation field
propagators in a background, being the inverse of
  $\Gamma^{(2,0)}[0,A]$, are required on the right-hand side in order to
extract the flow of $\Gamma_k[a=0,A]$ on the left-hand side. In the following
we will concentrate on determining these inverse propagators
in the background gauge with gauge parameter $\alpha=0$, i.e., in the
so-called Landau-DeWitt gauge. These are directly related to a
covariantization of the well-studied (inverse) Landau-gauge
propagators $\Gamma_{\mathrm{Landau}}^{(2)}(p^2)$, which is a
well-suited gauge for functional RG calculations as it is a fixed
point of the RG flow \cite{Litim:1998qi}.  {Moreover, FRG flows
  for correlation functions in the Landau-DeWitt gauge directly relate to
  those obtained within fully gauge-invariant flows for the
  geometrical effective action
  \cite{Branchina:2003ek,Pawlowski:2003sk}. It is interesting to note
  that in such a setting the difference between background and
  fluctuation correlation functions is controlled by Nielsen
  identities \cite{Pawlowski:2003sk}, and indeed supports the above
  relation between Landau-gauge propagators and background gauge
  propagators.}

The relation between the inverse propagators in {these} two
gauges can be parameterized by
\begin{equation}
\Gamma_k^{(2,0)}[0, A] = \Gamma_{k,\text{Landau}}^{(2)}(\mathcal{D}[A])+F_{\mu\nu} f_{\mu\nu}(D),
\label{eq:cov}
\end{equation} 
where $\mathcal{D}[A]$ is a suitable background covariant differential
  operator, reducing to the Laplacian for vanishing background field. The
covariant derivative with respect to the background field is given by
$D_{\mu}^{ab}= \partial_{\mu}\delta^{ab}+ g f^{abc}A_{\mu}^c$.  The function
$f_{\mu\nu}$ occurring in combination with the Yang-Mills field strength
$F_{\mu\nu}$ in the second term cannot be determined from the knowledge of the
Landau-gauge propagators alone. It is nonsingular for vanishing argument in
order to guarantee the proper Landau-gauge limit. For backgrounds with
vanishing field strength (such as a pure Polyakov loop as studied in
\cite{Braun:2007bx}), the background propagators are exactly determined.

In this work, we will particularly be interested in backgrounds of nonzero
field strengths. For this, we approximate the $f$ terms by a minimal
reconstruction in terms of replacing the dependence on the  momentum  $p^2$ of the
Landau-gauge propagators by the corresponding background-covariant Laplacians
for the transversal, longitudinal and ghost modes, $p^2\to \mathcal{D}_{\text
  T}, \mathcal{D}_{\text L}, \mathcal{D}_{\text{gh}}$, respectively, 
\begin{equation}
\mathcal{D}_{T\, \mu \nu}= -D^2\delta_{\mu \nu}+ 2 i
g\, F_{\mu \nu}, \quad \mathcal{D}_{L\, \mu \nu}= - D_{\mu}D_{\nu}, 
\end{equation}
and $ \mathcal{D}_{\text {gh}}=-D^2$. Here the transversal Laplacian also contains
the spin-1 coupling to the background field.
This leads to the following construction
of the background field inverse propagators
\begin{equation}\label{eq:Rk}
\Gamma_{k\,\mathrm{Landau}}^{(2)}(p^2) \to \Gamma_k^{(2,0)}[0, A]=
\Gamma_{k\,\mathrm{Landau}}^{(2)}(\mathcal{D}) ,
\end{equation}
where $\mathcal{D}= \mathcal{D}_{\text{T,L,gh}}$, respectively. 

For the flow equation \eqref{eq:flow}, also the regulator needs to be
specified. Background-gauge invariance requires a dependence of
  the regulator on the background-covariant momentum. RG invariance for
  non-vanishing IR cut-off, \cite{Pawlowski:2005xe,Pawlowski:2001df},
  and spectral considerations \cite{Gies:2002af}, then lead to
\begin{equation}
R_k= \Gamma_k^{(2,0)}(k^2)\, r(y), \quad y= \frac{\mathcal{D}}{k^2}. 
\label{eq:optreg}
\end{equation}
The regulator shape function $r(y)$ encodes the detailed prescription of
the momentum-shell integration and will be chosen as $r(y)= e^{-y}$ here. 

{It has been shown that the form of the regulator can be used to
maximize the physics content within a given approximation
\cite{Litim:2000ci,Pawlowski:2005xe}, very similarly to the
construction of improved and perfect actions on the lattice. For the
numerical computations in the present paper we use results for Landau
gauge propagators \cite{Fischer:2008uz,jan}, that have been obtained
within functional optimization \cite{Pawlowski:2005xe}. For the
general considerations we leave the shape function $r(y)$ in
\eq{eq:Rk} unspecified.}

{Finally, it should be emphasized that the use of background field flows
  is not limited to the pure Yang-Mills sector, but has also successfully been
  applied in QCD calculations including quark fluctuations
  \cite{Gies:2002hq,Braun:2005uj,Braun:2008pi,Braun:2009gm}.}

\section{Effective-action flow in a selfdual background}\label{selfdual}

We are interested in computing the flow of ${\Gamma}_k$ in a derivative
expansion, keeping all operators of the type $(F_{\mu\nu}^a F_{\mu\nu}^a)^n$,
with $n\in\mathbbm{N}$. This can be summarized in terms of an effective
potential $W({F^2})$ which is an analytic function of ${F^2=
F_{\mu\nu}^a F_{\mu\nu}^a}$. For this computation, it suffices to evaluate the
flow equation in a background of covariantly constant field strength that allows for a unique identification
of $(F_{\mu\nu}^a F_{\mu\nu}^a)^n$ operators.

To understand why the flow is independent of the specific choice of
background, consider theory space, which is the space of all couplings of
operators of a given field content, compatible with the chosen symmetries. The
flow equation defines a vector field in this space by determining a $\beta$
function for each running coupling. After a basis is chosen in this space,
i.e. the operators are specified uniquely as functions of the field content, a
specific field configuration will only serve to distinguish different
operators. The $\beta$ function of a specific running coupling can be
evaluated by projecting the flow equation onto a suitable field configuration,
but it will not depend on this configuration as long as it uniquely determines
the desired operator.

In principle, a covariantly constant colormagnetic field would be sufficient
to project onto the effective potential of the field strength
invariant. However, the background has a second meaning in the present
formalism: the flow is regularized with respect to the Laplace-type spectra in
the given background, see \Eqref{eq:optreg}. Purely magnetic backgrounds now
suffer from the problem of the tachyonic Nielsen-Olesen mode in the spectrum
of fluctuations \cite{Nielsen:1978rm} typically spoiling perturbative
computations \cite{Dittrich:1980nh}. The tachyonic mode indicates the 
instability of the colormagnetic background, which may be quantified by an imaginary 
part of the effective action as a decay rate. The functional RG can indeed deal with
this tachyonic mode problem owing to its well-controlled IR regularization
\cite{Gies:2002af,Braun:2005uj}. Still, the technical complications are
substantial and the physics of the tachyonic mode in terms of, e.g., decay
rates of constant magnetic backgrounds is not of primary interest.

Instead, we propose to use the only known stable constant background
field, a covariantly-constant selfdual background with $F_{\mu\nu}^a
\equiv \widetilde{F}_{\mu\nu}^a$, as first analyzed in
\cite{Minkowski:1981ma} and
\cite{Leutwyler:1980ma,Leutwyler:1980ev}. The covariant Laplacian
$\mathcal{D_{\text{T}}}$ has nontrivial zero modes $a_0$
(``chromons''), as discussed in detail in
  \cite{Leutwyler:1980ma,Leutwyler:1980ev}, which carry important
perturbative contributions, such as a significant contribution to the
1-loop $\beta$ function. As a disadvantage, the selfdual background
will not uniquely project onto the desired operators, since
$F_{\mu\nu}^a F_{\mu\nu}^a$ and $F_{\mu\nu}^a
\widetilde{F}_{\mu\nu}^a$ become indistinguishable. Whereas this can
potentially lead to systematic errors in the determination of the
effective potential $W({F^2})$, our important conclusions drawn from
the $\beta$ function of the running coupling will be unaffected, see
below.

The covariantly constant selfdual background is given by the gauge potential
\begin{equation}
A_\mu^a = - \frac{1}{2} F_{\mu\nu} x_\nu n^a, \quad n^a=\text{const.}, \quad
n^2=1, 
\end{equation}
where
\begin{equation}
F_{12}=F_{34}\equiv f=\text{const.}
\end{equation}
Apart from antisymmetric partner components, all other components are zero and
the field strength is of abelian type as the commutator of two gauge
potentials vanishes. Here, $n^a$ is a constant vector in color space that can
be rotated into the Cartan subalgebra.  For later use, we define $\nu_l$ (with
$l=1,...,N_c^2-1$) as the eigenvalues of $n^a (T^a)^{bc}$ and $f_l = \vert
\nu_l \vert f$.  For the traces, we need the spectral properties of the
relevant operators $\mathcal{D}_{\text{T}},-D^2$ in 4 dimensions, which
can be found in appendix \ref{spectra}.

In the following, we treat the zero mode $a_{0\mu}$, satisfying
$\big(\mathcal{D}_{\text{T}}\big)_{\mu\nu}^{ab} a_{0\nu}^b=0$, separately, as 
it requires special attention for the construction of a suitable regulator.
The gauge field is decomposed according to
\begin{equation}
  \mathcal{A}_\mu = A_\mu + a_{\text{T}\mu} + a_{\text{L}\mu} = A_\mu +
  a'_{\text{T}\mu} + a_{\text{L}\mu} +a_{0\mu}, 
\end{equation}
where $A_\mu$ is the selfdual background field, $a_{\text {T,L}}$ are
transversal and longitudinal gluon modes, and $a_{\text T}'$ denotes the
 transversal modes except for the zero mode. 
Our truncation can then be written as
\begin{eqnarray}
&&\Gamma_k[a,A]\nonumber\\
&=&\int d^4x \Bigl[ \frac{1}{2} a_{\text T \mu}^{a}
  \big(\Gamma_{k,\text T}^{(2,0)}(\mathcal{D}_{\text{T}})\big)_{\mu\nu}^{ab}\,
  a_{\text T \nu}^{b}+ \bar c^a
  \big(\Gamma_{k,\text{gh}}^{(2,0)} \big)^{ab} c^b \nonumber\\
&{}&+ \frac{1}{2}a_{\text L \mu}^a{}
  \big(\Gamma_{k,\text L}^{(2,0)}(-D^2)\big)_{\mu\nu}^{ab}\,
  a_{\text L \nu}^b  \Bigr],
\end{eqnarray}
where $c,\bar c$ denote the ghost fields. 

The inverse propagators for gluons and ghosts are taken from Landau-gauge
calculations. In the deep infrared, a family of solutions exists that is
parameterized by the infrared boundary condition for the ghost propagator
\cite{Fischer:2008uz, Maas:2009se}. They can be written as
\begin{eqnarray}\label{eq:invprops} 
\Gamma_{k \rightarrow 0, A}^{(2,0)}[0,0](p^2)&=&p^2 Z_A(p^2) \Pi_{\rm T} (p) \id, \nonumber\\
\Gamma_{k \rightarrow 0,c}^{(2,0)}[0,0](p^2)&=&p^2 Z_c(p^2)\id.
\end{eqnarray} 
The transversal projector satisfies $\Pi_{\rm T\, \mu \nu}(p)=\delta_{\mu
  \nu}-\frac{p_{\mu}p_{\nu}}{p^2}$, and the identity is understood to apply to
color indices.  Herein the wave function renormalizations
\begin{eqnarray}\label{eq:wfren}
Z_A(p^2\to 0)\simeq (p^2)^{\kappa_A}\,,\quad
Z_c(p^2\to 0)\simeq (p^2)^{\kappa_c}
\end{eqnarray} 
are parameterized by critical exponents $\kappa_A$ and $\kappa_c$ in the deep infrared.
In the Landau gauge, the longitudinal propagator remains trivial as the longitudinal
modes decouple completely from the flow.

The so-called \emph{decoupling solution}, found in lattice studies of
Landau-gauge Yang-Mills theory \cite{lattice}, and also in
  continuum studies, \cite{Fischer:2008uz,Cornwall:1981zr}, is characterized by the
critical exponents
\begin{equation}
  \kappa_A=-1\,, \quad {\rm and} \quad 
  \kappa_c=0\,,\label{eq:decoupling}
\end{equation}
which corresponds to a positivity-violating gluon propagator
\cite{Fischer:2008uz,Cucchieri:2004mf}, indicating the confinement of
gluons.

\begin{figure*}[t]
  \hspace*{0.0cm}
\includegraphics[width=0.41\linewidth]{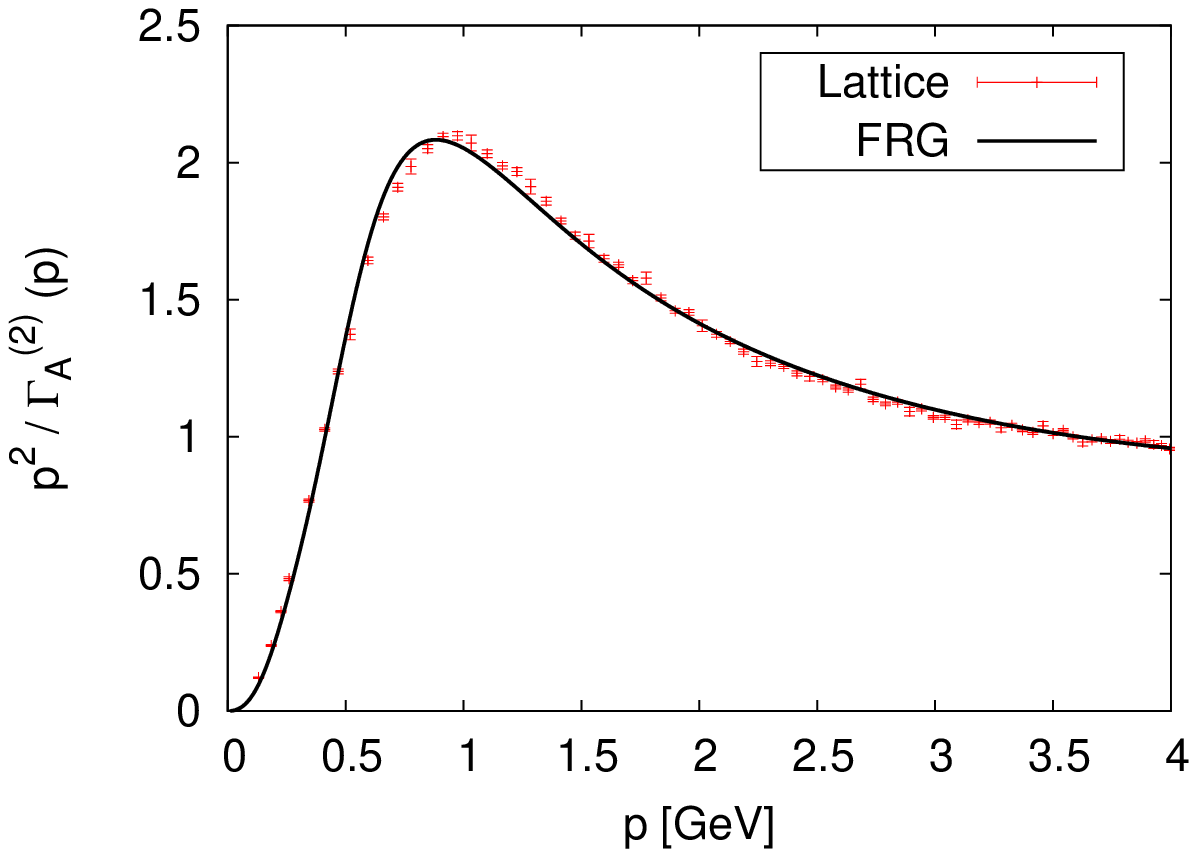}
  \hspace*{2.5cm}
\includegraphics[width=0.41\linewidth]{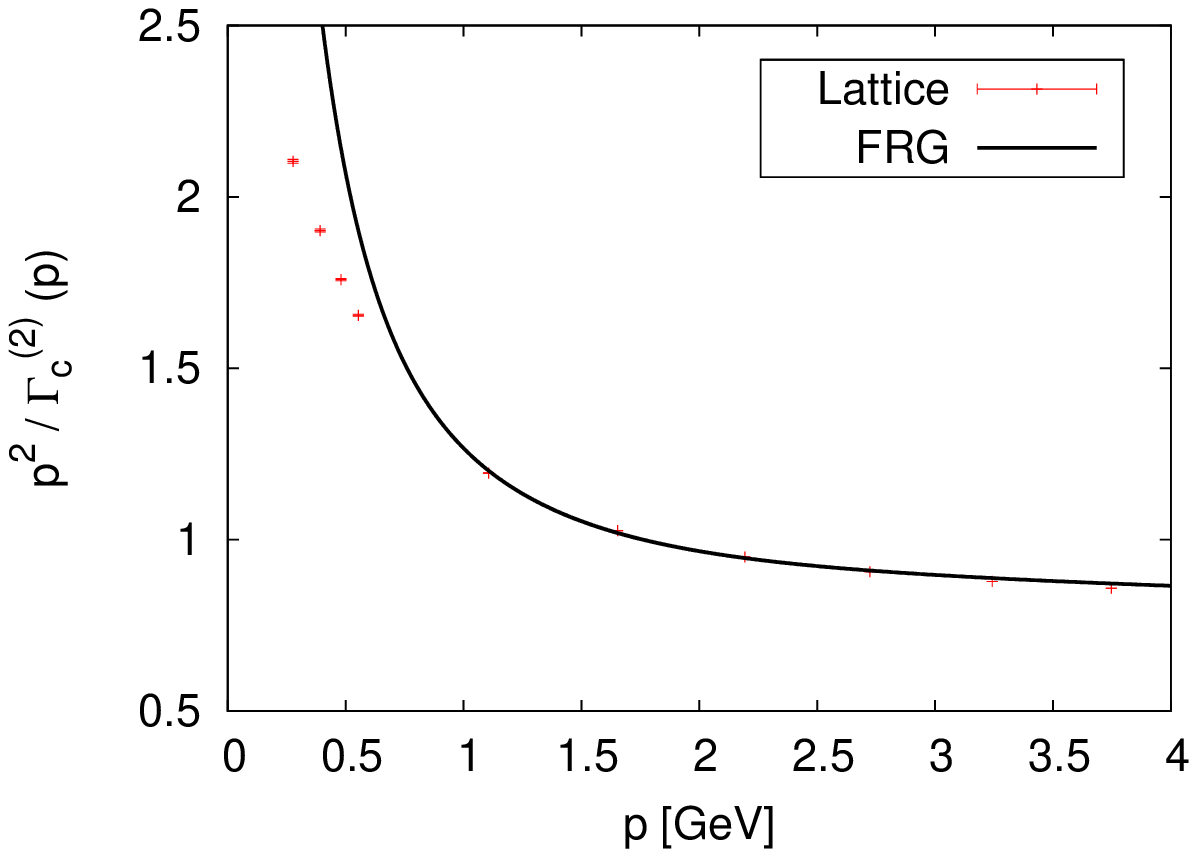} 
\caption{{Momentum dependence of the gluon (left panel) and ghost (right panel) 
2-point functions. We show the FRG results from 
Ref.~\cite{Fischer:2008uz} (black solid line) and from lattice simulations from 
Ref.~\cite{lattice} (red points).}}
\label{fig:propagators}
\end{figure*}

The \emph{scaling solution}, first identified in \cite{von
  Smekal:1997is}, has only one independent critical
exponent $\kappa_c$, as the
non-renormalization theorem for the ghost-gluon vertex \cite{Taylor:1971ff}
(proven to all-orders in perturbation theory) implies the sum rule
\begin{equation}
 \kappa_A = -2 \kappa_c \label{eq:sumrule}
\end{equation}
in $d=4$ dimensions \cite{Lerche:2002ep,Fischer:2006vf}. In most functional
computations we are led to ($d=4$)
\begin{eqnarray}\label{eq:scaling}  
\kappa_c=0.59535...\quad {\rm and} \quad \kappa_A=-2\kappa_c=- 1.1907..., 
\end{eqnarray} 
being the value for the optimized regulator \cite{Pawlowski:2003hq}.  The
regulator dependence in functional RG computations leads to a $\kappa_c$ range of
$\kappa_c\in [0.539\,,\, 0.595]$, see \cite{Pawlowski:2003hq}; for a
specific flow, see \cite{Fischer:2004uk}. In
Fig.~\ref{fig:propagators}, we show the momentum dependence of the ghost- and
gluon propagator as obtained from a functional RG study~\cite{Fischer:2008uz}
in comparison to lattice results \cite{lattice}. 

The scaling solution satisfies both, the Kugo-Ojima confinement criterion
$\kappa_c>0$ featuring an infrared enhanced ghost \cite{Kugo:1979gm}, as well as the
Gribov-Zwanziger scenario \cite{Gribov:1977wm,Zwanziger:1993dh}. The latter relates color
confinement to horizon conditions for gluons and ghosts which for the scaling
solution is satisfied if
\begin{equation}
 \kappa_c > \frac{1}{2}.
\end{equation}

The gauge correlation functions can also be used to relate color
  confinement to quark confinement: as shown in \cite{Braun:2007bx}, the
  infrared behavior of the propagators determines the effective potential of a
  Polyakov-loop order parameter for the deconfinement phase transition. 
  From the behavior of the effective potential, a sufficient criterion
    for quark confinement can be deduced, implying
\begin{equation}
 \kappa_c > \frac{1}{4}
\end{equation}
for the scaling solution. The decoupling solution also satisfies a
corresponding confinement criterion. Hence both the decoupling and the
scaling solution for the propagators correspond to scenarios where all
color charges are confined. The scaling solution, however, is the only
solution which is comaptible with global BRST invariance, see
\cite{Fischer:2008uz,vonSmekal:2008es}. For further work on the
  infrared behaviour of Landau-gauge Yang-Mills theory see, e.g.,
  \cite{Zwanziger:2009je}.

{The asymptotic forms of the propagators can be
parameterized as
\begin{equation}
\Gamma_{k,A}^{(2,0)}(p^2) =\Gamma^{(2)}_{k,A}(p^2)  \Pi_{\rm T} (p) \id , \quad
\Gamma_{k,c}^{(2,0)}(p^2)=\Gamma^{(2)}_{k,c}(p^2)\id,
\end{equation}
where the scalar functions $\Gamma^{(2)}_{k,A/c}(p^2)$ are the IR-regularized
generalizations of \Eqref{eq:invprops}  \cite{jan}, 
\begin{eqnarray}
\Gamma_{k,A}^{(2)}(p^2) &=& \gamma_{A} \frac{(p^2 + c_A
  k^2)^{1+\kappa_A}}{(\Lambda_{\text {QCD}}^2)^{\kappa_A}}, \nonumber\\
\Gamma_{k,c}^{{(2)}}(p^2) &=& \gamma_{c} \frac{p^2(p^2 + c_c k^2)^{\kappa_c}}{(\Lambda_{\text QCD}^2)^{\kappa_c}}. \label{eq:props}
\end{eqnarray}
}%
Here, the $\gamma_{A,c}$'s are simple proportionality constants which account
for the difference between $\Lambda_{\text{{QCD}}}$ and the scale where this
asymptotic form takes over.  The constants $c_{A,c}$ are manifestations of the
regulator dependence in the asymptotic regime; generically, they are of order
$\mathcal O(1)$. In the absence of any IR regularization, i.e., $k\to0$
  or $c_{A,c}\to 0$, Eq.~\eqref{eq:props} reduces to the standard form,
  cf. Eqs.~\eqref{eq:invprops}, \eqref{eq:wfren}. 

For the regularization of the zero modes, some care is required, since the
seeming standard choice $\mathcal{D}=\mathcal{D}_{\text{T}}$ would imply that the
regulator function $r(y)$ is always evaluated at the singular point
$y=\mathcal{D}/k^2 \to 0$ (because $\mathcal{D}_{\text{T}}=0$ on the zero-mode
subspace). In particular, this would not properly account for the decoupling
of the zero mode, once it acquires a field-dependent mass, which already happens perturbatively \cite{Leutwyler:1980ma}. Instead, we choose
$\mathcal{D}=-D^2$ as the argument of the regulator on the zero-mode subspace which
makes the regulator satisfy all standard requirements. On the zero-mode
subspace, we have $\mathcal{D}=-D^2\to 2f_l$, cf. first line of \Eqref{eq:spec} for
$n=m=0$. 

The flow equation for the background potential in the selfdual background can
finally be written as
\begin{widetext}
\begin{eqnarray}
\pat \Gk[A]&=& \frac{3}{2} \tr \,\partial_t R_{k, A}(-D^2)\left(\Gamma_{k, A}^{(2)}+R_{k, A}(-D^2) \right)^{-1}
- \tr \,\partial_t R_{k, c}(-D^2)\left(\Gamma_{k,c}^{(2)}+R_{k, c}(-D^2) \right)^{-1}\nonumber\\
&{}&+\frac{1}{2} \tr\, \partial_t R_{k, L}(-D^2) \left(\Gamma_{k, L}^{(2)}+ R_{k, L}(-D^2) \right)^{-1}
 + \frac{1}{2} \tr\, P_0 \partial_t R_{k, A}(-D^2)\left(\Gamma_{k, A}^{(2)}+R_{k, A}(-D^2) \right)^{-1}
, \label{eq:Gint}
\end{eqnarray}
\end{widetext}
where $P_0$ denotes the projector onto the zero-mode subspace. 
The zero-mode trace simply corresponds to the terms $m=n=0$ in the spectrum
\eqref{eq:spec}. The subscripts $A,c$ and $L$ in the regulator function imply
the use of the appropriate transversal, ghost or longitudinal two-point
function in its argument.

The trace measure factor in \Eqref{eq:Gint} corresponds to the density of
states in a selfdual background field, given by $[f_l/(2\pi)]^2$
\cite{Leutwyler:1980ma}.

\section{relation between the $\beta$ function and the critical exponents}\label{betafunction}

In the following we will investigate the asymptotic form of the $\beta$
function of the background running coupling, defined as the prefactor
of the $F_{\mu\nu}^a F_{\mu\nu}^a$ term:
\begin{equation}
\Gamma_k[A] = \int d^4 x \, \frac{1}{4 g^2} F_{\mu\nu}^a F_{\mu\nu}^a + \dots
\quad \to \quad \Gamma[f]=\Omega\,  \frac{f^2}{g^2},\label{coupling}
\end{equation}
where $\Omega$ denotes the spacetime volume. For asymptotic freedom, the
Gau\ss{}ian fixed point of the $\beta$ function has to be UV attractive
(i.e. IR repulsive).  On the other hand,  a minimum requirement for the
confinement of color charges is an interacting theory in the long-range
limit. This requirement is incompatible with an IR attractive Gau\ss ian fixed
point. As we show below, this seemingly elementary condition imposes a new
confinement criterion on the  exponents $\kappa_A$ and $\kappa_c$.

According to \Eqref{coupling} the $\beta$ function can be evaluated from the
first term in a Taylor expansion of the effective action in powers of the
background field
\begin{equation}
\beta_{g^2} := \pat g^2 = -g^4\, \pat \frac{1}{g^2} = -\frac{g^4}{\Omega}\, \pat
\Gk[A]\Big|_{f^2}.\label{eq:beta}
\end{equation}
Here $|_{f^2}$ denotes the projection onto the Taylor coefficient at
order $f^2$. Although there can be ambiguities in the projection onto
$(F_{\mu \nu}F_{\mu \nu})^n$ and $(F_{\mu \nu}\widetilde{F}_{\mu \nu})^n$
for a selfdual background, this does not affect the $\beta$ function,
as the second operator is only non-zero for even powers $n$ due to parity
conservation.
 
For the evaluation of \Eq{eq:beta}, some care is needed as the trace
  over the Laplace-type spectra and the projection onto the $f^2$ order do not
  commute. As the degeneracy factor in the trace already carries a factor of 
$\frac{f^2}{(2 \pi)^2}$, all $f$ dependence outside the operator trace can
already be ignored due to the projection, yielding
\begin{widetext}
\begin{eqnarray}
\pat \Gk[A]\Big|_{f^2}
&=& \Biggl[\Biggl(\frac{3}{2} \frac{\pat \Gamma_{k\,
    A}^{(2)}(k^2)}{\Gamma_{k\, A}^{(2)}(k^2)} \left(\frac{\Gamma_{k\,
      A}^{(2)}(0)}{\Gamma_{k\,A}^{(2)}(k^2)}+1 \right)^{-1}
- \frac{\pat \Gamma_{k,c}^{(2)}(k^2)}{\Gamma_{k, c}^{(2)}(k^2)} 
\left(\frac{\Gamma_{k, c}^{(2)}(0)}{\Gamma_{k,c}^{(2)}(k^2)}+1 \right)^{-1}+1 \Biggr) \tr\,e^{-\frac{-D^2}{k^2}}\nonumber\\
&{}& +  \frac{\pat \Gamma_{k\, A}^{(2)}(k^2)}{\Gamma_{k\, A}^{(2)}(k^2)}\left(\frac{\Gamma_{k\, A}^{(2)}(0)}{\Gamma_{k\,A}^{(2)}(k^2)}+1 \right)^{-1}\tr\, e^{-\frac{2f_l}{k^2}}\Biggr] \Bigg|_{f^2}
. \label{eq:betaflow}
\end{eqnarray}
\end{widetext}
To obtain the $\beta$ function, we extract the coefficient of the
expansion of the heat-kernel trace over coordinate and color space at
second order in $f$,
\begin{eqnarray}
\Tr_{x\text{c}} e^{- \left(\frac{-D^2}{k^2} \right)} &=&
\frac{\Omega}{(4\pi)^2}\,  \sum_{l=1}^{\Nc^2 -1} \,
\frac{f_l^2}{\sinh^2 \left( \frac{f_l}{k^2}
  \right)}\nonumber\\
&=&\frac{\Omega}{(4 \pi)^2}\sum_l\left(1-\frac{f_l^2}{3} +\mathcal{O}(f^4)\right).
\end{eqnarray}
Evaluating ${\partial_t \Gamma^{(2)}_{k, A/c}}$ with the infrared asymptotic form of the inverse
propagator \Eqref{eq:props} yields the infrared form of the $\beta$ function,
whereas using the perturbative form of the inverse propagators
$\Gamma^{(2)}(p^2) \sim p^2$ results in the standard one-loop form of
the $\beta$ function. Recovering this universal term within our setting
can be viewed as a simple check of our formalism.

Focusing on the infrared asymptotic form of the propagators
\Eqref{eq:props}, we obtain to lowest order in the background-field
  coupling 
\begin{equation}
\beta_{g^2} = - \frac{\Nc g^4}{(4 \pi)^2}
\Biggl(
-\frac{(1+\kappa_A)}{c_{\kappa}+1}+\frac{2}{3}(1+\kappa_c)-\frac{1}{3}+8
\frac{1+\kappa_A}{c_{\kappa}+1}\Biggr),
\label{eq:betaIR} 
\end{equation}
where we have used that $\sum_l^{\Nc^2 -1} \nu_l^2 =\Nc$. Here we have defined
\begin{eqnarray}
c_{\kappa} = \frac{\Gamma_{k\, A}^{(2)}(0)}{\Gamma_{k\, A}^{(2)}(k^2)}=
 \left(\frac{c_A}{1+c_A} \right)^{1+\kappa_A}. 
\end{eqnarray}
Incidentally, the perturbative 1-loop $\beta$ function $\beta_{g^2}=-\frac{22}{3}
\frac{\Nc g^4}{(4\pi)^2}$ is obtained from \Eq{eq:betaIR} in the limit of
perturbative critical exponents $\kappa_{A,c}\to 0$ and in the absence of an
additional IR regularization $c_{A,\kappa}\to 0$. 

For retaining an interacting theory in the long-range limit, the $\beta$
  function near the Gau\ss ian fixed point must not change sign. 
This gives us a new bound on the IR values of the $\kappa$'s,
\begin{equation}
 -\frac{(1+\kappa_A)}{c_{\kappa}+1}+\frac{2}{3}(1+\kappa_c)-\frac{1}{3}+8 \frac{1+\kappa_A}{c_{\kappa}+1} > 0, \label{eq:b1}
\end{equation}
For the accidental case of an equality, the sign of the $\beta$
function would have to be determined from higher order terms in $g^2$
and a corresponding additional bound would follow from the prefactor of
  these higher-order terms.

\begin{figure}[!here]
 \includegraphics[width=0.45\linewidth]{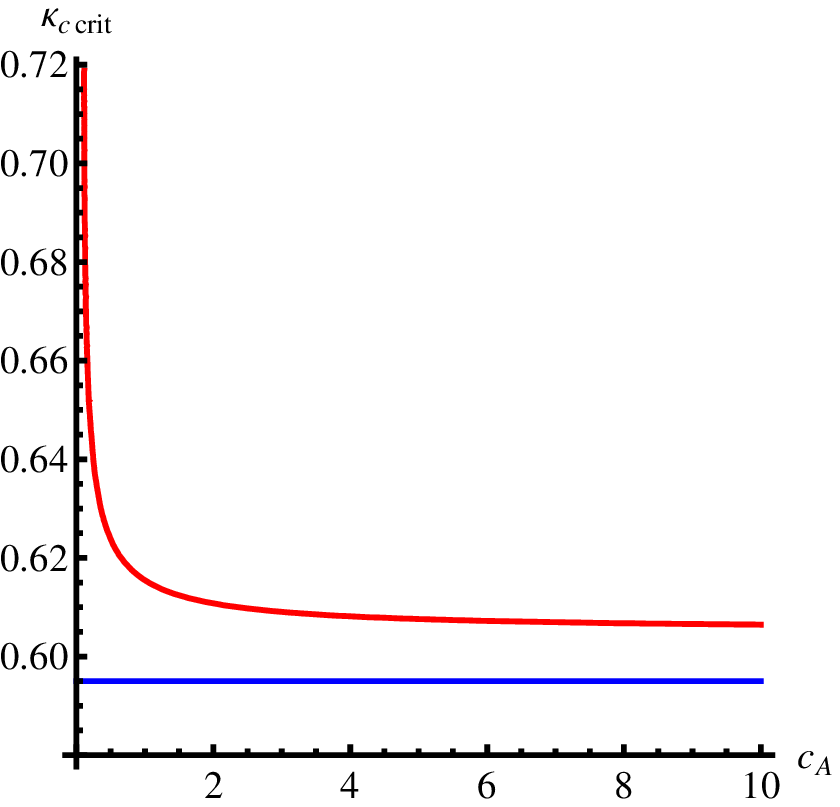} 
\includegraphics[width=0.45\linewidth]{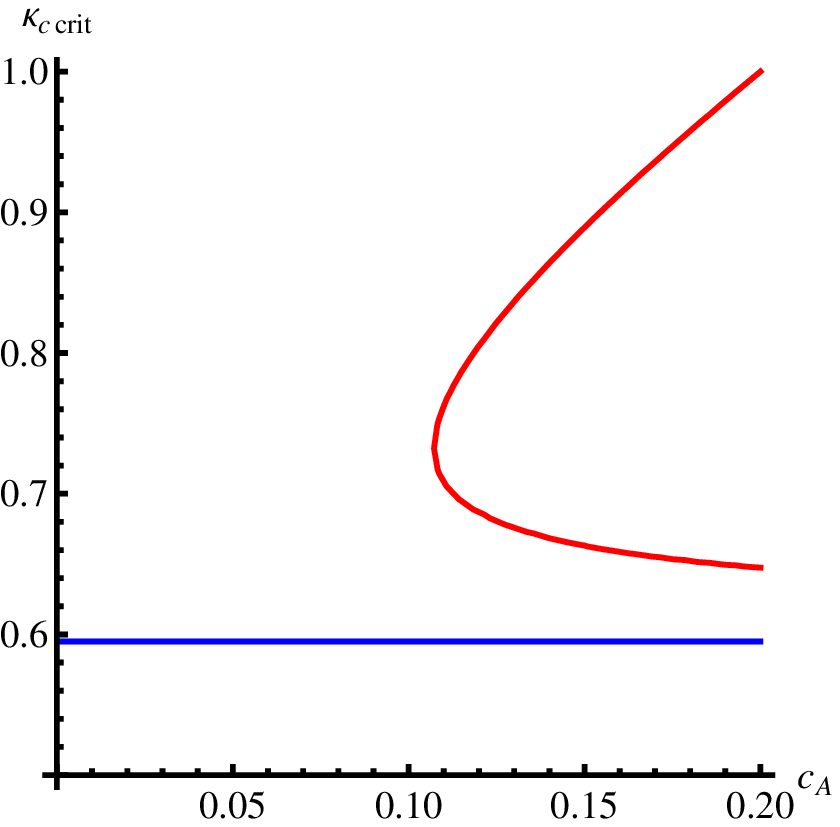}
\caption{$\kappa_{c\, \rm crit}$ as a function of $c_A$. Left panel:
    continuous branch of $\kappa_{c, \text{crit}}$ that exists for all $c_A$
    (all regularization schemes); for the $c_A$ values displayed here,
    $c_A\gtrsim\mathcal{O}(1)$, $\kappa_{c, \text{crit}}$ represents an upper
    bound for the ghost critical exponent $\kappa_c$. Right panel: for
    $0.1\lesssim c_A \lesssim 0.2$, the bound bifurcates, such that all
    $\kappa_c$ values to the left of the {red/upper} curve are allowed. The horizontal
  line marks the value $\kappa_c\simeq0.595$ as found in must functional
  calculations for the scaling solution which obviously satisfies our
  confinement criterion.}
\label{kappacrit}
\end{figure}

Let us first concentrate on the scaling solution: Using the sum rule
$\kappa_A = -2 \kappa_c$, this bound yields a critical value of
$\kappa_{c, \rm crit}(c_A)$ as a function of the regulator-dependent constant
$c_A$.  For an interacting IR theory, the true critical exponent has to
  satisfy
\begin{equation}
\kappa_c< \kappa_{c, \text{crit}}(c_A),
\label{eq:bound}
\end{equation}
such that our criterion imposes for the first time a relevant upper bound on
the Landau-gauge exponents. The strongest bound (minimum upper bound)
is achieved in the limit of $c_A \rightarrow \infty$, where
\begin{equation}
 \kappa_{c, \rm crit}(c_A \rightarrow \infty) = \frac{23}{38} \simeq 0.6053.
\end{equation}
The maximum upper limit for $\kappa_{c, \rm crit}$ is reached for $c_A
\approx 0.1073$, where $\kappa_{c, \rm crit}\simeq 0.72767$. For
  completeness let us mention that, for values of $c_A\lesssim 0.2$, the inequality \Eqref{eq:b1} can also be satisfied if
$\kappa_{c}> \kappa_{c, {\rm crit}\, 2}$, as the nonlinear inequality
  \Eq{eq:b1} bifurcates. We find that allowed values for $\kappa_c$ lie to
the left of the {red/upper} curve (see right panel of
Fig.~\ref{kappacrit}). For $0\leq c_A\lesssim 0.1073$, only the bound
  $\kappa_c<1$ from unitarity \cite{Lerche:2002ep} remains. However, let us
  stress that the limit $c_A\to0$ corresponds to a highly asymmetric
  regularization as the contribution from transverse gluons to the $\beta$
  function is removed in this limit (note that $c_\kappa\to \infty$ for
  $c_A\to 0$ and $\kappa_A<-1$).  For this case, the Gau\ss{}ian fixed point
is naturally IR repulsive for all values of $\kappa_c > \frac{1}{2}$.

To summarize, our argument based on  the mere existence of an
  interacting IR regime imposes a bound on the IR critical exponent
  $\kappa_c<\kappa_{c, \text{crit}}$. Both sides of this inequality are
  regulator dependent, such that the bound has to be satisfied in every
  regularization scheme as a necessary criterion. A sufficient criterion is
  therefore given by imposing that $\kappa_c<\kappa_{c,
    \text{crit,min}}=23/38\simeq0.6053$ holds for all regularization schemes. 

This bound is furthermore required to be satisfied due to the
following argument: an IR fixed point for the background-field running
coupling is observed in the background-field gauge in a derivative
expansion \cite{Gies:2002af} as well as for the running coupling fixed
at the ghost-gluon vertex in the Landau gauge in a vertex expansion
\cite{von
  Smekal:1997is,Lerche:2002ep,Fischer:2006vf}. Both running couplings
are defined with the aid of nonrenormalization theorems arising from
gauge invariance.  From the conjecture that both running couplings are
actually linked on all scales, consistency requires the existence of
the fixed point in both gauges.  Even though such a fixed point is not
observed in the simplified truncation leading to \Eqref{eq:beta}, it
is expected to arise in a larger truncation as used in
\cite{Gies:2002af}. But already without knowing the full IR flow of the
  background field coupling, consistency of the  leading-order result of
  \Eqref{eq:betaIR} with the Landau-gauge coupling gives rise to the bound on
  the scaling exponent. This consistency would be spoiled if the above
first coefficient of the $\beta$ function changed sign. Therefore,
we infer that $\kappa_{c}< \frac{23}{38}$ is also required for the
consistency between the background-gauge and the Landau-gauge results,
  independently of higher-order terms that generate an IR fixed 
  point of the coupling.

Physically our upper bound for $\kappa_c$ may be interpreted as a
criterion for an IR interacting theory and thus ultimately for
confinement. Therefore, we conclude that for $\kappa_{c}$ larger than
some critical value Yang-Mills theory would not be confining.
Therefore it is very reassuring that all results for $\kappa_c$
  obtained by functional methods satisfy the sufficient criterion
  bound $\kappa_c < \frac{23}{38}$ , c.f. \Eq{eq:scaling}. 

It is important to note that also that the decoupling solution for the
propagators $\kappa_A =-1$, $\kappa_c=0$ that has been found in many
lattice simulations (and can possibly be understood as a different way
to deal with the gauge ambiguity in the Landau gauge, see
\cite{Maas:2009se}) fulfills our confinement criterion which in this
case is given by the more general bound \Eqref{eq:b1}.

\section{Gluon condensation from the effective potential}\label{effpot}

Dimensional transmutation and the scale anomaly going along with  the
quantization of Yang-Mills theory require a non-zero expectation value for
the energy-momentum tensor. This implies a non-trivial vacuum structure:
\begin{equation}
 \langle F^2 \rangle \neq 0.
\end{equation}
Indeed phenomenological estimates indicate a non-zero value
\cite{Shifman:1978bx}, being interpreted as a condensation of gluons in
the vacuum. These findings have been corroborated by other methods
\cite{Andreev:2007vn,Rakow:2005yn}, as well as lattice gauge theory \cite{Di
  Giacomo:1981wt}.

The effective action for a colormagnetic background field has been evaluated
at one-loop order in \cite{Savvidy:1977as}, where a non-trivial minimum has
been found (Savvidy vacuum). Due to the tachyonic mode of the gluon propagator
on such a background this configuration is unstable \cite{Nielsen:1978rm},
which can be amended by introducing a spatial inhomogeneity into the field
configuration. This vacuum configuration is referred to as the Copenhagen
vacuum \cite{Nielsen:1978rm,Nielsen:1978tr,Ambjorn:1978ff,Ambjorn:1980ms}. A
more severe problem is posed by the fact that the minimum lies in the
non-perturbative domain, i.e. beyond the perturbative Landau pole of the
Yang-Mills coupling, questioning the relevance of a perturbative estimate.

First non-perturbative studies from functional methods indeed provided 
indications for gluon condensation \cite{Reuter:1994zn}.

Here we will further pursue this question, based on the knowledge of
  full correlation functions. As a simple parametrization of these correlation
  functions, we still use the asymptotic form displayed in \Eq{eq:props} which
  we amend with the $k$-dependent critical exponents $\kappa_A(k)$ and
  $\kappa_c(k)$ in accordance with the propagators in \cite{jan}. A suitable
  interpolation between $\kappa_{A,c}(k\to\infty)\to 0$ and the corresponding
  IR values $\kappa_{A,c}(k\to0)\to \kappa_{A,c}$ of \Eq{eq:scaling} can
  parameterize the full momentum dependence of the correlation functions.
This allows to evaluate the effective potential from an integrated form of the
flow equation:
\begin{eqnarray}
W_k({F^2})&=& -\frac{1}{\Omega}\int_0^{k_{\rm UV}} \frac{dk}{k} \frac{1}{2}{\rm STr} \partial_t R_k \left(\Gamma_k^{(2)}+R_k \right)^{-1}\nonumber\\
&{}& + W_{k_{\rm UV}}({F^2})\label{intflow},
\end{eqnarray}
where $k_{\rm UV}$ is an initial ultraviolet scale, and ${F^2=
  F_{\mu\nu}^a F_{\mu\nu}^a= 4f^2}$.  The initial condition
$W_{k_{\rm UV}}({F^2})$ is fixed deep inside the perturbative regime at
$k_{\rm UV} = 10$ GeV, 
\begin{equation}
W_{\rm UV}=  W_{k=10 \rm GeV} = 
{ \frac{F^2}{ 16 \pi\, 0.2294},}
\end{equation}
{where we have used the peak of the maximum of the gluon dressing function
  $1/Z_A(p^2)$, {see left panel of Fig.~\ref{fig:propagators},} for
  relating our YM scales to that used in lattice computations.  The
  normalization is such that the related string tension $\sigma$ (computed on
  the lattice) is given by $\sqrt{\sigma}=440$ MeV. Our initial condition is
  thus self-consistently determined from our main input, the Landau-gauge
  ghost and gluon propagators, {which goes along with a coupling
    $\alpha_{\text{S}}(k=10\text{GeV})\simeq 0.2294$ at the initial scale.  At
    this} initial cut-off scale, $k=10$GeV,
  higher-order operators do not contribute significantly to the effective
  potential, which is determined exclusively by the functional form of the
  classical action.} Alternatively, an RG-improved initial condition in the
form of the one-loop effective potential could be used.

\begin{figure}[!here]
 \includegraphics[width=1\linewidth]{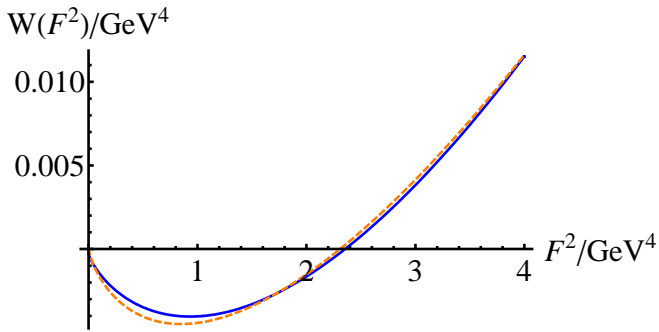}
\label{fit_pot}
\caption{The effective potential for SU(3) as a function of $F^2$
  (thick blue line), and the one-loop inspired fit to the numerical
  data of the form $a F^2 \, \ln b F^2$ (orange dashed line).}
\end{figure}

\Eqref{intflow} can then directly be integrated numerically. For SU(3),
we find a non-trivial minimum at 
\begin{equation}
 F^2 \approx  0.93 \,{\rm GeV}^4 = (3.46 \lqcd)^4, \label{eq:Fqvalue}
\end{equation}
where we have used $\lqcd= 284$ MeV, cf. solid line in Fig.~\ref{fit_pot}.

We emphasize that the formation of the non-trivial minimum is
  mainly driven by the dynamics in the mid-momentum regime, the form
  of the propagators in the deep infrared is not crucial here. The
existence of a gluon condensate in the vacuum thereby occurs for both
the scaling and the decoupling solution, and its value does not
  significantly depend on the infrared asymptotics.  This agrees with the
  observation in \cite{Braun:2007bx}, where quark confinement can also
  be inferred from the propagators, independent of their infrared
  asymptotics.

Note that on the right-hand side of the flow equation the color eigenvalues
$\vert \nu_l \vert$ enter. We may choose these along one of the two directions
of the Cartan subalgebra, which yields an uncertainty of about $10 \%$ in the
value of the minimum.

As a simple parametrization, we fit the effective potential to a
  function  of the form
\begin{equation}
 W(F^2)= a F^2 \,\ln (b F^2), \label{eq:logfit}
\end{equation}
which is inspired by the corresponding one-loop results
\cite{Savvidy:1977as} with two fit parameters $a$
and $b$, cf. dashed line in Fig.~\ref{fit_pot}). The fit yields $a=0.00528$ and
$b=0.433 \text{GeV}^{-4}$.

With some reservations, the one-loop result might thus be interpreted
  as providing indeed a reasonable qualitative estimate of the true
  nonpertubative functional form of the effective potential. Let us stress
  once more that the one-loop calculation is, of course, not reliable, as the
predicted minimum lies in a regime inaccessible to perturbation theory. It is
only its qualitative prediction of the functional form of the potential
which seems to be altered little by non-perturbative corrections.

A few comments on the numerical values for the gluon condensate are in order:
first of all, as the selfdual background does not distinguish between the
invariants $F^2$ and $F\tilde{F}$, our result for the condensate receives
contributions from both operators; under the assumption that condensates of
both types exist, the value given in \Eqref{eq:Fqvalue} should be considered
as an upper bound for the phenomenologically more relevant $F^2$
condensate. In this sense, our result $\langle F^2 \rangle/4\pi
  \simeq0.074 \text{GeV}^4$ is well compatible with recent
phenomenological estimates, $\langle F^2 \rangle/4\pi \simeq0.068(13)
  \text{GeV}^4$ from spectral sum rules \cite{Narison:2009vy} (note
  that our field definition differs from that of \cite{Narison:2009vy} by a
  rescaling with the coupling, cf. \Eqref{coupling}). The good
  agreement might even indicate that the contribution from $F\tilde{F}$
  operators is rather small. Also, we expect the
inclusion of dynamical quark degrees of freedom to decrease our condensate
value slightly, owing to their screening property. 

We have used a background that can only be considered as an approximation of
the true ground state locally. For studies involving a more realistic gauge-field configuration than the one considered here
 see \cite{Galilo:2010fn} and references therein.

Within the {\em leading-log} model which is based on the assumption that
  the infrared effective action is essentially given by its one-loop
  functional form \eqref{eq:logfit}, the condensate value is related to the
  string tension in a simple manner \cite{Adler:1981as}: the static potential between two opposite
  color charges arising from the nonlinear field equations following from
  \Eqref{eq:logfit} grows linearly with distance, $V=\sigma r$, where the
  string tension $\sigma$ and the minimum of the action obey, $\sqrt{\sigma} =
  (Q\kappa)^{1/4}$. Here, $Q=4/3$ is a simple color factor for SU(3), and
  ${\kappa=(1/4)F^2}|_{\text{min}}$ denotes the minimum of the action,
  i.e., the condensate value. From our result \eqref{eq:Fqvalue}, we obtain
  the estimate
\begin{equation}
\sigma^{1/2} \simeq 747 \text{MeV}
\end{equation}
which, in view of the rather restrictive truncation {and the simplicity of the
  leading-log model}, compares rather favorably with the {value $\sigma^{1/2}
  \simeq 440\text{MeV}$ which we used for fixing the scale of the propagators
  in the first place}. We should stress that the string-tension
  $\sigma$ appears in two very different meanings in our calculation: It
    first occurs as an input scale for fixing the initial condition for the flow
    equation, i.e., it fixes the absolute scale of the propagators. We then
  derive the \emph{physical} string tension in a nontrivial way from
  the minimum of the effective potential via the leading-log
    model. This output is linked to a mechanism of confinement, and has therefore
  acquired a physical meaning beyond pure scale fixing. With regard to
  the approximations involved in our calculation and, in particular, in the
  confinement model used to map our results onto the physical string tension,
  the order-of-magnitude equivalence of ``input scale'' and ``output
  physics'' is satisfactory.

\section{Conclusions}\label{conclusion}

Based on the knowledge of low-order gauge correlation functions from
  Landau-gauge calculations, we have used an approximate mapping onto
  propagators in the Landau-DeWitt gauge in a background field. This allowed us to
  extract nonperturbative information about the effective action of Yang-Mills
  theories. We have specialized to a selfdual constant background which --
  apart from technical simplicity -- is known to be stable against
  fluctuations and thus a candidate for at least local approximations of the
  Yang-Mills vacuum.

  We have concentrated on two properties of the effective action: the running
  of the background field coupling and the form of the effective potential for
  the self-dual field strength. From the simple requirement that the
  background field coupling should remain finite, and the theory thus
  interacting, in the long-range limit, we have derived a nontrivial criterion
  for the infrared behavior of Landau-gauge correlation functions. As
  interactions are mandatory for confinement, our criterion may be viewed as a
  confinement criterion. For the scaling solution of these correlation
  functions which is characterized by a single critical exponent $\kappa_c$,
  this criterion translates into a new upper bound on $\kappa_c$. In its
  sufficient version, the confinement criterion is satisfied if
\begin{equation}
  \kappa_c < \kappa_{c\,\rm crit, min}=\frac{23}{38} \simeq 0.6053,
\end{equation}
where $\kappa_{c, \text{crit,min}}$ is a minimum value obtained in a specific
regularization scheme. If $\kappa_c$ satisfies this bound in any scheme, the
system remains interacting in the infrared. In general, both sides of the
inequality are regularization-scheme dependent, such that the condition that
$\kappa_c[R_k]<\kappa_{c, \text{crit}}[R_k]$ is a necessary criterion for
confinement which then has to be satisfied for each regulator
$R_k$.

Read together with the confinement criterion for quark confinement as derived
from the Polyakov loop potential in
\cite{Braun:2007bx}, $\kappa_c>\frac{1}{4}$, and the horizon condition for
color confinement  \cite{Zwanziger:1993dh}, $\kappa_c>\frac{1}{2}$, our new
criterion defines a window for the infrared critical exponent,
$0.5<\kappa_c\lesssim0.6053$, thereby tightly fixing the possible infrared
behavior of the gauge correlation functions for the scaling solution in the
infrared. It is reassuring that the result $\kappa_c \simeq0.595$
obtained from many functional calculations lies precisely in this window.
 
Furthermore, we numerically investigate the full effective potential
for the field strength invariant ${ F^2=F_{\mu\nu}^a F_{\mu\nu}^a}$ and find a
non-trivial minimum at $ {F^2\simeq 0.93 \,{\rm GeV}^4}$. Our
calculation is fully non-perturbative, and therefore supports the
conclusions drawn from the one-loop effective action
\cite{Savvidy:1977as} in a non-trivial way.

Most importantly, effective models of confinement such as the {\em
    leading-log} model or dielectric confinement models receive strong support
  from our results. These models map the nontrivial vacuum structure onto
  classical field equations. For instance, it can be shown rather
  straightforwardly in the leading-log model that a nontrivial minimum of the
  effective action at nonzero ${F^2}$ is already sufficient to produce a
linearly rising potential for static color charges at long distances
\cite{Adler:1981as}. Our calculation therefore provides for an explicit
example how different pictures of confinement can not only exist in parallel but
actually support each other, potentially being two sides of the same coin. 
A similar observation has been made for the Coulomb gauge in the
  Hamiltonian approach and the Gribov-Zwanziger confinement scenario
  \cite{Reinhardt:2008ek}.

  Quantitatively, our estimate of the selfdual gluon condensate ${F^2\simeq
    0.93 \,{\rm GeV}^4}$ compares favorably with phenomenological
    estimates \cite{Narison:2009vy} (${F^2\simeq 0.85 \,{\rm GeV}^4}$). It is
    also in agreement with the large-distance limit of the static potential
  between a quark and antiquark, yielding a string tension of
  $\sigma^{1/2}\simeq 747\text{MeV}$ within the leading-log model of
  confinement \cite{Adler:1981as}. Our nonperturbative calculations, starting
  from the microscopic theory, thus can be interpreted as providing a
  fundamental justification of the two basic assumptions of the leading-log
  model: the functional form of the effective action \eqref{eq:logfit} and the
  free parameter given in terms of the minimum of the effective action, i.e.,
  the gluon condensate. Our studies therefore provide an explicit example of
  how seemingly different confinement scenarios, the Kugo-Ojima and
  Gribov-Zwanziger scenario on the one hand, and the leading-log model on the
  other hand, can not only exist in parallel but actually support each other.

Our results are affected by several sources of uncertainties: first of
  all, the mapping of Landau-gauge propagators onto background-field
  propagators is not unique, but has been realized in a minimal-coupling
  fashion. Whereas we have concentrated onto lowest-order correlation
  functions in the Landau gauge, also higher orders contribute to the full
  effective potential in the background gauge. In principle, it is
  straightforward to include such higher-order correlation functions, as,
  e.g., computed in \cite{Cucchieri:2008qm}, into our formalism. Also, differences
  between gluonic field invariants, such as $(F^2)^n$ and $(F \tilde F)^n$,
  which have been ignored for the selfdual background can be treated with
  suitable heat-kernel expansion techniques in future studies.

\acknowledgments Helpful discussions with J.~Braun are gratefully
acknowledged.  This work was supported by the DFG support under Gi 328/1-4 and
Gi 328/5-1 (Heisenberg program), through the DFG-Research Training Group
"Quantum- and Gravitational Fields" (GRK 1523/1), and by Helmholtz Alliance HA216/
EMMI. A.E. gratefully acknowledges the hospitality of Perimeter Institute, where this work was completed.

\begin{appendix}

\section{Spectral properties of Laplace-type operators}\label{spectra}

Let us summarize the spectral properties of the Laplace-type operators
involved. We concentrate here on the particularities of the selfdual
background. The general trace technology in covariantly constant backgrounds
as it is relevant for the computations in this paper can be found in
\cite{Reuter:1994zn,Gies:2002af}. The selfdual-background spectra needed in
this work are: 
\begin{eqnarray}
  \text{Spec}\Big\{ -D^2 \Big\} &=& 2 f_l (n+m+1), \quad n,m=0,1,2,\dots \nonumber\\
  \text{Spec}\Big\{ \mathcal{D}_{\text{T}} \Big\} &=& \left\{
    \begin{array}{lll}
      2 f_l (n+m+2)&,\,\,&\text{multiplicity}\quad 2\\
      2 f_l (n+m)&,\,\,&\text{multiplicity}\quad 2
    \end{array}
  \right.,\nonumber\\
&& \label{eq:spec}
\end{eqnarray}
where $f_l=|\nu_l| f$, and $\nu_l$ denotes the eigenvalues of the adjoint
color matrix $n^a T^a$. The covariant spin-1 Laplacian
$\mathcal{D}_{\text{T}}$ has a double zero mode for $n=m=0$ which is due to
the symmetry between colorelectric and colormagnetic field. Using the
trace technology of \cite{Reuter:1994zn,Gies:2002af}, the spectral problem of
the longitudinal Laplacian $\mathcal{D}_{\text{L}}$ can be mapped onto that of
$-D^2$, such that \Eqref{eq:spec} is sufficient for the calculation in the
main part of the paper.

Defining $\Tr'$ as the trace without the zero mode, we make the following
useful observation for the trace over some function $\mathcal F$:
\begin{widetext}
\begin{eqnarray}
\Tr_{x\text{cL}}' \mathcal F(\mathcal{D}_{\text{T}}) &=& 2
\sum_{l=1}^{\Nc^2-1}\, \left( \frac{f_l}{2\pi}\right)^2 \left\{
\sum_{n,m=0}^\infty \mathcal F \big( 2f_l(n+m+2)\big) + 
\sum_{n=0}^{\infty} \sum_{m=1}^{\infty} \mathcal F  \big( 2f_l(n+m)\big)
+ \sum_{n=1}^{\infty} \mathcal F  \big( 2f_l n\big)\right\} \nonumber\\
&=& 4
\sum_{l=1}^{\Nc^2-1}\, \left( \frac{f_l}{2\pi}\right)^2 
\sum_{n,m=0}^\infty \mathcal F \big( 2f_l(n+m+1)\big) \nonumber\\
&=& 4 \Tr_{x\text{c}} \mathcal F (-D^2), \label{eq:iso}
\end{eqnarray}
\end{widetext}
where the trace subscripts denote traces over coordinate space ``$x$'', color
space ``c'' and Lorentz indices ``L''.  In other words, there exists an
isospectrality relation between $-D^2$ and the non-zero eigenvalues of
$\mathcal {D}_{\text{T}}$. Similar isospectrality properties are also
  known for the Dirac operator in a self-dual homogeneous background
  \cite{Dunne:2002qf}. As a 
consequence, all gluon and ghost modes except for the two zero modes couple in
the same fashion to the selfdual background. This basic observation has been
used for the decomposition of terms in \Eqref{eq:Gint}.
\end{appendix}

\end{document}